\documentclass[prl,print£¬superscriptaddress,showpacs,onecolumn,showkeys]{revtex4}
\usepackage[dvips]{graphics,color}
\usepackage{amsfonts}
\usepackage{amsmath}
\usepackage{amssymb}
\usepackage{graphicx}
\usepackage{fancyhdr}

\setcitestyle{open={},close={}}

\input{tcilatex}

\begin{document}

\title{SPIN-PARITY EFFECT IN VIOLATION OF BELL'S INEQUALITIES FOR ENTANGLED
STATES OF PARALLEL POLARIZATION}
\author{HAIFENG ZHANG, JIANHUA WANG, ZHIGANG SONG }
\affiliation{Institute of Theoretical Physics and Department of Physics, Shanxi
University, Taiyuan, Shanxi 030006, China}
\author{J. -Q. LIANG}
\affiliation{{Institute of Theoretical Physics and Department of Physics, Shanxi U
niversity, Taiyuan, Shanxi 030006, China} \\
{*jqliang@sxu.edu.cn}}
\author{L. -F. WEI}
\affiliation{State Key Laboratory of Optoelectronic Materials and Technologies, School of
Physics and Engineering, Sun Yat-Sen University, Guangzhou 510275, China}
\affiliation{Quantum Optoelectronics Laboratory, School of Physics and Technology,
Southwest Jiaotong University, Chengdu 610031, China}
\received{ 15 September 2016}
\revised{ 1 December 2016}

\begin{abstract}
Bell inequalities (BIs) derived in terms of quantum probability statistics
are extended to general bipartite-entangled states of arbitrary spins with
parallel polarization. The original formula of Bell for the two-spin singlet
is slightly modified in the parallel configuration, while, the inequality
formulated by Clauser-Horne-Shimony-Holt remains not changed. The violation
of BIs indeed resulted from the quantum non-local correlation for spin-$1/2$
case. However, the inequalities are always satisfied for the spin-$1$
entangled states regardless of parallel or antiparallel polarizations of two
spins. The spin parity effect originally demonstrated with the antiparallel
spin-polarizations (Mod. Phys. Lett. B28, 145004) still exists for the
parallel case. The quantum non-locality does not lead to the violation for
integer spins due to the cancellation of non-local interference effects by
the quantum statistical-average. Again the violation of BIs seems a result
of the measurement induced nontrivial Berry-phase for half-integer spins.
\end{abstract}

\maketitle


%
%%%%%%%%%%%%%%%%%%%%%%%%%%%%%%%%%%%%%%%%%%%%%%%%%%%%%%%%%%%%%%%%%%%%

%\accepted{(Day Month Year)}
%\comby{(xxxxxxxxxx)}
\thispagestyle{fancy}

\part{\protect\large 1.\ Introduction}

Non-locality being the most peculiar characteristic of quantum mechanics
seems inconsistency with the classical field-theory based on the
relativistic causality. Quantum entanglement as one of the most striking
feature of quantum mechanics has become a resource for quantum computing and
quantum information with technological breakthroughs in these areas.{%
\textsuperscript{\cite{2,3,4}}} Bell's theorem known as Bell's inequality
(BI) proved that the existence of entangled quantum states has no classical
counterpart and therefore provides a possibility of quantitative test for
non-local correlations. BI was originally derived under classical statistics
with the assumption of "locality", which means that physical systems cannot
be instantly affected by distant objects in a space-like distance. Various
extensions of the original BI have investigated {\textsuperscript{%
\cite{26,5}}} and attracted considerable attentions both theoretically and
experimentally.{\textsuperscript{\cite{2,3,11,12,13,14,16}}} The violation
of BIs is predicted in some particular entangled-states with overwhelming
experimental evidence,{\textsuperscript{\cite{5,6,7,8,9,10,11,12,13,14}}}
which invalidates local realistic interpretations of quantum mechanics. The
profound application of non-locality is mainly in the quantum information
with a space--time separation not achievable for the classical systems. The
violation of BIs constitutes an ability to faithfully produce, control and
read out entangled states of qubit-pairs. The non-locality now has become a
powerful resource{\textsuperscript{\cite{3,16}}} of quantum information
science.{\textsuperscript{\cite{17,18,19,20,21,22,23,24,25}}} Nevertheless,
there are still many open problems in relation with both bipartite and
multipartite entanglements.{\ }Although the experimental evidence seems
strongly support the non-local nature the underlying physical-principle is
obscure{\textsuperscript{\cite{15}}} and many aspects in relation with the
initial debate remain to be fully understood. The non-determinism and
non-locality have been bearing continuously theoretical scrutiny ever since
the birth of quantum mechanics.

\textbf{\ }To understand underlying physical principle of the violation of
BIs\textbf{\ }we\textbf{\ }in a previous paper{\textsuperscript{\cite{1}} }%
adopted the density operators of entangled states to evaluate the
measurement-outcome-correlations in terms of quantum probability-statistics.
With the separation of density operators into local and non-local parts the
non-local outcome-correlations are obtained explicitly. The local part of
density operator describes the classical probability-statistics in the
absence of quantum interference between two components of the entangled
state. The BI is verified from the local part of density operator{%
\textsuperscript{\cite{27}}} alone with the assumption of
measurement-outcome-independence. On the other hand the violation of BIs
indeed resulted from non-local correlations of entangled states. A spin
parity effect in the violation of BIs was predicted{\textsuperscript{%
\cite{1}} }that BIs are violated by the half-integer spin entangled-states
but not the integer spins. Moreover the violation in entangled spin-states
is seen to be an effect of Berry phase (BP) induced by relative-reversal
measurements of two spins. We established for the first time a relation
between two kinds of non-localities, namely, the violation of BI violation
and the so-called dynamic non-locality regarded as the geometric phase
interference of quantum states.\textsuperscript{\cite{1}} The spin-parity
phenomenon was originally demonstrated in terms of a spin singlet following
the model of Bell. It is a natural question that whether or not the
spin-parity effect exists only for these particular states with antiparallel
spin-polarizations. We in the present paper consider bipartite-entangled
state with parallel spin-polarization, which leads to the modification of
original form of BI. However the inequality derived by
Clauser-Horne-Shimony-Holt (CHSH) remains not changed. The spin-parity
phenomenon for the violation of BIs seems independent of the particular form
of entangled states.

\part{\protect\large 2. Bell's inequalities for spin-1/2 and spin-1
entangled states of parallel spin-polarization and their violations}

\part{\protect\large 2.1 Spin-1/2 state}

The violation of BIs was theoretically verified long ago for spin-$1/2$
bipartite entanglement with arbitrary superposition of opposite
spin-polarization states\textsuperscript{\cite{33,34}} in stead of the spin
singlet in the original formulation of Bell. It was revisited in a recent
paper{\textsuperscript{\cite{1}}} based on quantum probability statistics,
which has advantage that both the BI and its violation can be formulated in
a unified manner with the help of state density-operator.{%
\textsuperscript{\cite{1}}} We consider in the present paper a two-spin
entangled state with parallel polarization such that%
\begin{equation}
|\psi \rangle =c_{+}|+,+\rangle +c_{-}|-,-\rangle  \label{1}
\end{equation}%
where $c_{\pm }$ are two arbitrary complex coefficients with the
normalization condition \TEXTsymbol{\vert}$c_{+}|^{2}+$\TEXTsymbol{\vert}$%
c_{-}|^{2}=1$ and $|\pm \rangle $ denote the usual spin-$1/2$ eigenstates ($%
\widehat{\sigma }_{z}|\pm \rangle =\pm |\pm \rangle $). Without losing
generality, the two arbitrary coefficients $c_{\pm }$ can be parameterized
as
\begin{equation*}
c_{+}=e^{i\eta }\cos \xi ,\qquad c_{-}=e^{-i\eta }\sin \xi ,
\end{equation*}%
with $\xi $ and $\eta $ being two real parameters. The density-operator $%
\overset{\wedge }{\rho }$ of entangled state Eq.(1) can be split into two
parts%
\begin{equation*}
\overset{\wedge }{\rho }=\overset{\wedge }{\rho }_{lc}+\overset{\wedge }{%
\rho }_{nlc}
\end{equation*}%
in order to see the non-local quantum correlation explicitly\textbf{.} Here,
the first part is a density operator of the complete mixed-state
\begin{equation*}
\overset{\wedge }{\rho }_{lc}=\cos ^{2}\xi |+,+\rangle \langle +,+|+\sin
^{2}\xi |-,-\rangle \langle -,-|,
\end{equation*}%
which describes two particles obeying the local or classical statistics in
the absence of quantum interference at all. While, the quantum interference
term
\begin{equation*}
\overset{\wedge }{\rho }_{nlc}=e^{2i\eta }\sin \xi \cos \xi |+,+\rangle
\langle -,-|+e^{-2i\eta }\sin \xi \cos \xi |-,-\rangle \langle +,+|
\end{equation*}%
denotes the non-local correlation, which remains even if the two particles
are separated in a space-like interval. Following Bell, two spins are
measured independently along arbitrary directions respectively, say ${%
\mathbf{a}}$ and ${\mathbf{b}}$. According to the quantum measurement
principle, measuring outcomes fall into the eigenvalues of projection
spin-operators $\hat{\sigma}\cdot {\mathbf{a}}$ and $\hat{\sigma}\cdot {%
\mathbf{b}}$, i.e. %
%\begin{equation}
$\hat{\sigma}\cdot \mathbf{a|\pm a}\rangle =\pm \mathbf{|\pm a}\rangle $,
and $\hat{\sigma}\cdot \mathbf{b|\pm b}\rangle =\pm \mathbf{|\pm b}\rangle $
%\end{equation}%
. Two orthogonal eigenstates of a projection spin-operator $\hat{\sigma}%
\cdot \mathbf{r}$ can be found explicitly as
\begin{equation*}
\mathbf{|+r}\rangle \mathbf{=}\cos (\frac{\theta _{r}}{2})|+\rangle +\sin (%
\frac{\theta _{r}}{2})e^{i\phi _{r}}|-\rangle
\end{equation*}%
and
\begin{equation*}
\mathbf{|-r}\rangle \mathbf{=}\sin (\frac{\theta _{r}}{2})|+\rangle -\cos (%
\frac{\theta _{r}}{2})e^{i\phi _{r}}|-\rangle
\end{equation*}%
which are called the spin coherent states of north- and south- pole gauges.$%
\textsuperscript{\cite{30}}$ Here, $\mathbf{r=(}\sin \theta _{r}\cos \phi
_{r},\sin \theta _{r}\sin \phi _{r}$,$\cos \theta _{r})$ with $\mathbf{r}=%
\mathbf{a},\mathbf{b}$ is an unit vector parameterized by the polar and
azimuthal angles ($\theta _{r}$ and $\phi _{r}$) in the coordinate frame
with $z$-axis along the direction of the initial spin-polarization.
Outcome-independent base vectors for two-particle measurements are seen to
be the direct product of eigenstates for operators $\hat{\sigma}\cdot
\mathbf{a}$ and $\hat{\sigma}\cdot \mathbf{b}.$ We can arbitrarily label
these four base vectors as{\textsuperscript{\cite{1}}}
\begin{equation}
|1\rangle \!\!=\!\!\mathbf{|\!\!+\!a,\!\!+\!b}\rangle ,\!|2\rangle \!\!=\!\!%
\mathbf{|\!\!+\!a,\!\!-\!b}\rangle ,\!|3\rangle \!\!=\!\!\mathbf{%
|\!\!-\!a,\!+\!b}\rangle ,|4\rangle \!\!=\!\!\mathbf{|\!\!-\!a,\!\!-\!b}%
\rangle .  \label{2}
\end{equation}%
Then the correlation probability for independent measurements of two spins
is the quantum statistical average of the correlation operator
\begin{equation*}
\hat{\Omega}(ab)=(\hat{\sigma}\cdot \mathbf{a)}(\hat{\sigma}\cdot \mathbf{b)}
\end{equation*}%
in the state $\overset{\wedge }{\rho }$
\begin{equation}
P(ab)=Tr[\hat{\Omega}(ab)(\hat{\rho}_{lc}+\hat{\rho}%
_{nlc})]=P_{lc}(ab)+P_{nlc}(ab),  \label{3}
\end{equation}%
Notice that the non-vanishing matrix elements of correlation operator in the
outcome-independent base vectors Eq.(2) are obviously%
\begin{equation*}
\Omega _{11}(ab)=\Omega _{44}(ab)=1
\end{equation*}%
and%
\begin{equation*}
\Omega _{22}(ab)=\Omega _{33}(ab)=-1,
\end{equation*}%
the correlation probability is found as
\begin{equation*}
P(ab)=\rho _{11}+\rho _{44}-\rho _{22}-\rho _{33}.
\end{equation*}%
The density-matrix elements denoted by $\rho _{ij}=\langle i|\overset{\wedge
}{\rho }|j\rangle $, ($i,j=1,2,3,4$) can be split into two parts
\begin{equation*}
\rho _{ii}=\rho _{ii}^{lc}+\rho _{ii}^{nlc},
\end{equation*}%
with local elements given by
\begin{equation*}
\rho _{11}^{lc}=\cos ^{2}(\xi )K_{a}^{2}K_{b}^{2}+\sin ^{2}(\xi )\Gamma
_{a}^{2}\Gamma _{b}^{2}
\end{equation*}%
\begin{equation*}
\rho _{22}^{lc}=\cos ^{2}(\xi )K_{a}^{2}\Gamma _{b}^{2}+\sin ^{2}(\xi
)\Gamma _{a}^{2}K_{b}^{2}
\end{equation*}%
\begin{equation*}
\rho _{33}^{lc}=\cos ^{2}(\xi )\Gamma _{a}^{2}K_{b}^{2}+\sin ^{2}(\xi
)K_{a}^{2}\Gamma _{b}^{2}
\end{equation*}%
\begin{equation*}
\rho _{44}^{lc}=\cos ^{2}(\xi )\Gamma _{a}^{2}\Gamma _{b}^{2}+\sin ^{2}(\xi
)K_{a}^{2}K_{b}^{2}.
\end{equation*}%
Where $K_{r}^{m}=\cos ^{m}(\theta _{r}/2)$ and $\Gamma _{r}^{m}=\sin
^{m}(\theta _{r}/2)$ with $r=a,b,c,d$ denoting measurement directions and $m$
the integer power-index throughout the paper. The non-local elements $\rho
_{11}^{nlc}$, $\rho _{44}^{nlc}$ have equal value but opposite signs with
respect to $\rho _{22}^{nlc}$, $\rho _{33}^{nlc}$

\begin{eqnarray}
\rho _{11}^{nlc} &=&\rho _{44}^{nlc}=-\rho _{22}^{nlc}=-\rho _{33}^{nlc}
\label{4} \\
&=&\frac{1}{2}\sin \xi \cos \xi \sin \theta _{a}\sin \theta _{b}\cos \left(
\phi _{a}+\phi _{b}-2\eta \right) .  \notag
\end{eqnarray}%
This property plays a crucial role, which we will see, in the violation of
BIs. The measurement outcome correlation has the same form
\begin{equation}
P_{lc}(ab)=Tr[\hat{\Omega}(ab)\hat{\rho}_{lc}]=\cos \theta _{a}\cos \theta
_{b},
\end{equation}%
but a positive sign different from that for the spin singlet with opposite
spin-polarizations{\textsuperscript{\cite{1}}} where the correlation formula
has a negative sign seen from Eq.(4) of Ref.(\cite{1}). This simple
sign-difference leads to a necessary modification of the original formula of
BI. In other words, the specific form of BI is state dependent. From the
local correlation Eq.(5) it is easy to find the modified BI being
\begin{equation}
|P_{lc}(ab)-P_{lc}(ac)|\leq 1-P_{lc}(bc),  \label{6}
\end{equation}%
different from the original{\textsuperscript{\cite{1}}} BI by a sign change
in front of the local correlation $P_{lc}(bc)$ on the greater side of
inequality. As a matter of fact, substitution of the correlation Eq.(5) into
the less side of BI Eq.(6) yields%
\begin{equation*}
|P_{lc}(ab)-P_{lc}(ac)|\leq |\cos \theta _{b}-\cos \theta _{c}|.
\end{equation*}%
Therefor it is straightforward to prove the modified BI in the parallel
spin-polarization
\begin{equation*}
|P_{lc}(ab)-P_{lc}(ac)|+P_{lc}(bc)\leq |\cos \theta _{b}-\cos \theta
_{c}|+\cos \theta _{b}\cos \theta _{c}\leq 1.
\end{equation*}%
The modified BI can be also verified by means of classical statistics
following Bell as shown in Appendix. The non-local correlation
\begin{equation*}
P_{nlc}(ab)=2\sin \xi \cos \xi \sin \theta _{a}\sin \theta _{b}\cos (\phi
_{a}-\phi _{b}-2\eta ),
\end{equation*}%
which comes from the quantum interference of coherent superposition of
two-particle states, is responsible for the violation of BI. To see the
violation of BI explicitly we assume that the parameters of superposition
coefficients are $\xi =\pi /4$ and $\eta =n\pi $ with $n$ being zero or
integer. The total measurement correlation including the non-local parts
becomes the well known quantum correlation-probability being a scalar
product of the two unit-vectors $\mathbf{a}$ and $\mathbf{b}$
\begin{equation}
P(ab)=\mathbf{a}\cdot \mathbf{b},  \label{7}
\end{equation}%
which has a sign difference compared with the opposite spin-polarization.{%
\textsuperscript{\cite{1}}} The violation of BI has been investigated
extensively{\textsuperscript{\cite{1}}} in terms of the quantum
correlation-probability$.$The CHSH inequality remains not changed, since a
common sign-difference of the measurement outcome correlation Eq.(5) for any
two directions does not affect the absolute value%
\begin{equation*}
P_{\mathrm{CHSH}}^{lc}=|P_{lc}(ab)+P_{lc}(ac)+P_{lc}(db)-P_{lc}(dc)|.
\end{equation*}%
We still have\textbf{\ }
\begin{equation*}
P_{CHSH}^{lc}\leq 2.
\end{equation*}%
With the quantum correlation-probability Eq.(7) it is also obviously to have
the well known formula
\begin{equation*}
P_{\mathrm{CHSH}}=|P(ab)+P(ac)+P(db)-P(dc)|\leq 2\sqrt{2}.
\end{equation*}

\part{\protect\large 2.2 Spin-1 state}

For the spin-$1$ entangled state of parallel spin-polarizations
\begin{equation*}
|\psi \rangle =c_{+}|+1,+1\rangle +c_{-}|-1,-1\rangle ,
\end{equation*}%
the spin-coherent states of spin-$1$ projection operator $\hat{s}\cdot
\mathbf{a}$ are found as$\textsuperscript{\cite{1}}$%
\begin{equation*}
|+\mathbf{a}\rangle _{1}=K_{a}^{2}|+1\rangle +\frac{1}{\sqrt{2}}e^{i\phi
_{a}}\sin \theta _{a}|0\rangle +\Gamma _{a}^{2}e^{i2\phi _{a}}|-1\rangle
\end{equation*}%
and
\begin{equation*}
|-\mathbf{a}\rangle _{1}=\Gamma _{a}^{2}|+1\rangle -\frac{1}{\sqrt{2}}%
e^{i\phi _{a}}\sin \theta _{a}|0\rangle +K_{a}^{2}e^{i2\phi _{a}}|-1\rangle ,
\end{equation*}%
where $\hat{s}_{z}|\pm 1\rangle =\pm |\pm 1\rangle $. The
measurement-correlation probability of two spin-$1$ particles initially
prepared in the entangled state with parallel spin-polarization can be
obtained by the local matrix elements
\begin{eqnarray*}
\rho _{(1)11}^{lc} &=&\cos ^{2}(\xi )K_{a}^{4}K_{b}^{4}+\sin ^{2}(\xi
)\Gamma _{a}^{4}\Gamma _{b}^{4}, \\
\rho _{(1)44}^{lc} &=&\cos ^{2}(\xi )\Gamma _{a}^{4}\Gamma _{b}^{4}+\sin
^{2}(\xi )K_{a}^{4}K_{b}^{4}, \\
\rho _{(1)22}^{lc} &=&\cos ^{2}(\xi )K_{a}^{4}\Gamma _{b}^{4}+\sin ^{2}(\xi
)\Gamma _{a}^{4}K_{b}^{4}, \\
\rho _{(1)33}^{lc} &=&\cos ^{2}(\xi )\Gamma _{a}^{4}K_{b}^{4}+\sin ^{2}(\xi
)K_{a}^{4}\Gamma _{b}^{4}.
\end{eqnarray*}%
However the four non-local matrix elements are all equal
\begin{equation}
\rho _{(1)22}^{nlc}=\rho _{(1)33}^{nlc}=\rho _{(1)11}^{nlc}=\rho
_{(1)44}^{nlc}=\frac{1}{8}\cos \xi \sin \xi \sin ^{2}\theta _{a}\sin
^{2}\theta _{b}\cos 2(\phi _{a}-\phi _{b}-\eta ),  \label{8}
\end{equation}%
the same as in the case of antiparallel spin-polarizations.%
\textsuperscript{\cite{1}} It was explained in our previous paper%
\textsuperscript{\cite{1}} that the minus sign difference between $\rho
_{11}^{nlc}$,$\rho _{44}^{nlc}$ and $\rho _{22}^{nlc}$,$\rho _{33}^{nlc}$ in
Eq.(4) for the spin-$1/2$ is actually a nontrivial BP resulted from the
relative reversal of two-spin measurements ($\rho _{22}^{nlc},\rho
_{33}^{nlc}$). In the spin-$1$ case the BP factor is only a trivial one $%
e^{i2\pi }=1$. Therefore, the contributions of non-local interference
between the same and opposite spin-polarization measurements cancel each
other. The total outcome correlation is originated from the local density
operator $\hat{\rho}_{(1)}^{lc}$ only, i.e.,
\begin{equation*}
P_{(1)}(ab)=P_{(1)}^{lc}(ab)=\cos \theta _{a}\cos \theta _{b}.
\end{equation*}%
There is no room for the violation of BIs in agreement with the previous
observation for the spin-$1$ entangled state with antiparallel spin
polarizations.\textsuperscript{\cite{1}}\emph{\ }It is interesting to see a
fact that the original form of BI is valid only for entangled states of
antiparallel spin-polarizations. However our observation, that BIs are
violated by the spin-$1/2$ but not the spin-$1$ entangled states, is true
regardless of antiparallel or parallel polarizations. Now we extend our
theorem to arbitrarily high spins.

\part{\protect\large 3. Spin-parity phenomenon }

For two spin-$s$ particles the entangled macroscopic quantum-state (MQS)
with parallel spin-polarization is defined as
\begin{equation}
|\psi \rangle =c_{+}|+s,+s\rangle +c_{-}|-s,-s\rangle .  \label{9}
\end{equation}%
The density operator of it can also be separated into the local part
\begin{equation*}
\hat{\rho}_{(s)}^{lc}=\cos ^{2}\xi |+s,+s\rangle \langle +s,+s|+\sin ^{2}\xi
|-s,-s\rangle \langle -s,-s|,
\end{equation*}%
obeying the local or classical theory and the non-local part
\begin{eqnarray*}
\hat{\rho}_{(s)}^{_{nlc}} &=&e^{2i\eta }\cos \xi \sin \xi |+s,+s\rangle
\langle -s,-s| \\
&&+e^{-2i\eta }\cos \xi \sin \xi |-s,-s\rangle \langle +s,+s|,
\end{eqnarray*}%
respectively. Here, the extreme state $|\pm s\rangle $ ( $\hat{s}_{z}|\pm
s\rangle =\pm s|\pm s\rangle $ ) is called the MQS wherein the minimum
uncertainty relation $|\langle \widehat{s}_{z}\rangle |=2\langle (\Delta
\widehat{s}_{x})^{2}\rangle ^{1/2}\langle (\Delta \widehat{s}%
_{y})^{2}\rangle ^{1/2}$ is satisfied. So that the state $|\psi \rangle $
defined in Eq.(9) is called the Bell cat, which is actually entangled Schr%
\H{o}dinger cat-states for both "dead" and both "life" cats. We assume that
the measurements are restricted on MQS, namely the spin coherent states $%
\mathbf{|\pm a}\rangle _{s}$ with $\hat{s}\cdot \mathbf{a|\pm a}\rangle
_{s}=\pm s\mathbf{|\pm a}\rangle _{s}$. These spin coherent states can be
generated from the extreme states $|\pm s\rangle $ such that%
\begin{equation*}
\mathbf{|\pm a}\rangle _{s}=\hat{R}|\pm s\rangle
\end{equation*}%
with the generation operator
\begin{equation*}
\hat{R}=e^{i\theta _{a}\mathbf{m}\cdot \hat{s}}.
\end{equation*}%
The unit-vector $\mathbf{m}$ in the $x-y$ plane is perpendicular to the
plane expanded by $z$-axis and the unit vector $\mathbf{a}$.

The explicit forms of spin coherent-states in the representation of Dicke
states are given by$\textsuperscript{\cite{30,31}}$
\begin{equation*}
|+\mathbf{a}\rangle _{s}=\sum_{m=-s}^{s}\binom{2s}{s+m}\!\!^{\frac{1}{2}%
}K_{a}^{s+m}\Gamma _{a}^{s-m}e^{i(s-m)\phi _{a}}|m\rangle ,
\end{equation*}%
\begin{equation*}
|-\mathbf{a}\rangle _{s}=\sum_{m=-s}^{s}\binom{2s}{s+m}^{\frac{1}{2}%
}K_{a}^{s-m}\Gamma _{a}^{s+m}e^{i(s-m)(\phi _{a}+\pi )}|m\rangle .
\end{equation*}%
In the outcome-independent base vector of two-particle measurements
(corresponding to Eq. (2), however, with the spin-$1/2$ replaced by $s$, )
density-matrix elements for the Bell cat-state are given by:%
\begin{eqnarray}
\rho _{11}^{lc} &=&\cos ^{2}(\xi )K_{a}^{4s}K_{b}^{4s}+\sin ^{2}(\xi )\Gamma
_{a}^{4s}\Gamma _{b}^{4s},  \notag \\
\rho _{22}^{lc} &=&\cos ^{2}(\xi )K_{a}^{4s}\Gamma _{b}^{4s}+\sin ^{2}(\xi
)\Gamma _{a}^{4s}K_{b}^{4s},  \notag \\
\rho _{33}^{lc} &=&\cos ^{2}(\xi )\Gamma _{a}^{4s}K_{b}^{4s}+\sin ^{2}(\xi
)K_{a}^{4s}\Gamma _{b}^{4s},  \notag \\
\rho _{44}^{lc} &=&\cos ^{2}(\xi )\Gamma _{a}^{4s}\Gamma _{b}^{4s}+\sin
^{2}(\xi )K_{a}^{4s}K_{b}^{4s},  \label{10}
\end{eqnarray}%
for the local part. With the above density-matrix elements Eq.(10) the
normalized (the correlation per spin value) local correlations become%
\begin{equation}
P_{(s)}^{lc}(ab)=(K_{a}^{4s}-\Gamma _{a}^{4s})(K_{b}^{4s}-\Gamma _{b}^{4s}),
\label{11}
\end{equation}%
which is also different from the opposite spin-polarization{%
\textsuperscript{\cite{1}}} by a negative sign compared with the Eq.(7) in
Ref.(\cite{1}). Using the corresponding local correlations Eq.(11) and
notice $K_{\alpha }^{4s}-\Gamma _{\alpha }^{4s}\leq 1$ for $\alpha =a,b,c,d$%
, it is obviously that CHSH inequality remains the same
\begin{equation*}
\left\vert
P_{(s)}^{lc}(ab)+P_{(s)}^{lc}(ac)+P_{(s)}^{lc}(db)-P_{(s)}^{lc}(dc)\right%
\vert \leq 2.
\end{equation*}%
While the BI is also modified as%
\begin{equation*}
1-P_{(s)}^{lc}(bc)\geq |P_{(s)}^{lc}(ab)-P_{(s)}^{lc}(ac)|.
\end{equation*}%
\textbf{\ }The non-local density-matrix elements are evaluated by

\begin{equation*}
\rho _{11}^{nlc}=\rho _{44}^{nlc}=2\sin \xi \cos (\xi )K_{a}^{2s}\Gamma
_{a}^{2s}K_{b}^{2s}\Gamma _{b}^{2s}\cos \left[ 2s\left( \phi _{a}+\phi
_{b}-2\eta \right) \right] ,
\end{equation*}%
for the measurements along the same spin-polarization direction. The
density-matrix elements for the measurements on opposite directions

\begin{equation}
\rho _{(s)22}^{nlc}=\rho _{(s)33}^{nlc}=(-1)^{2s}\rho _{(s)11}^{nlc},
\label{12}
\end{equation}%
possess an additional geometric phase factor $e^{i2s\pi }=(-1)^{2s}$
compared with $\rho _{(s)11}^{nlc}=\rho _{(s)44}^{nlc}$. It was well
explained in the previous paper{\textsuperscript{\cite{1}}} that the
geometric phase or BP resulted from the relative reversal of
spin-polarization measurements in the following matrix-element evaluation
\begin{eqnarray}
\rho _{_{(s)}22}^{nlc} &=&e^{2i\eta }\sin \xi \cos (\xi )\langle +s|+\mathbf{%
a}\rangle \langle +\mathbf{a}|+s\rangle \langle -s|-\mathbf{b}\rangle
\langle -\mathbf{b}|-s\rangle  \notag \\
&&+e^{-2i\eta }\sin \xi \cos (\xi )\langle -s|+\mathbf{a}\rangle \langle +%
\mathbf{a}|-s\rangle \langle +s|-\mathbf{b}\rangle \langle -\mathbf{b}%
|+s\rangle .  \label{13}
\end{eqnarray}%
$\break $

Remarkably we come to the same conclusion as in the case of antiparallel
spin-polarizations{\textsuperscript{\cite{1}}} that the non-local outcome
correlation vanishes $P_{(s)}^{nlc}(a,b)=0$ in the integer-spin Bell
cat-state, since the BP factor is trivial and thus the four elements become
equal\textbf{\ (}$\rho _{(s)11}^{nlc}=\rho _{(s)22}^{nlc}$) seen from
Eq.(12). However, in the half-integer\textbf{\ }spin\textbf{\ }case an
additional minus sign of the BP factor leads to\textbf{\ }$\rho
_{(s)11}^{nlc}=-\rho _{(s)22}^{nlc}$, and the non-local part of correlation
probability\textbf{\ }$P_{(s)}^{nlc}(ab)$\textbf{\ }does not vanish any more%
\textbf{.} The spin parity phenomenon discovered in the previous paper{%
\textsuperscript{\cite{1}}} still exists for the Bell cat-states with
parallel spin-polarization (both "dead" and both "life" cats)\textit{\ : }
the BI can be violated for the Bell cat-states of half-integer spins but not
for the states of integer spins.

\part{\protect\large 4. Conclusions and Discussions}

The proposed formulation of quantum probability-statistics with
state-density operator has advantage to separate the local and non-local
measurement correlations. The non-local part, which comes from the quantum
interference between two superposed state-components, can be evaluated
independently to see why and how the violation of BIs takes place. For the
bipartite-entanglement states of parallel spin-polarization the local
measurement-correlations have only a sign difference with respect to the
opposite spin-polarizations demonstrated in Ref.(\cite{1}). The original BI
is slightly modified due to the sign difference, however, CHSH inequality is
not changed in the entangled state of parallel spin-polarization. The
previous observation of spin-parity phenomenon{\textsuperscript{\cite{1}}}
still holds, that the BIs are indeed violated for entangled MQS of
half-integer spins but not the integer-spins. The violation for the
half-integer spins can be understood as the effect of geometric phase
induced by the relative reversal of spin measurements. We establish a
relation between two-type of non-localities, namely, the violation of BI and
the dynamic non-locality resulted from the geometrical phase.

Although the specific form of original BI depends on the initial state the
spin-parity effect seems state independent. Our generic arguments of the
spin-parity effect particularly for the nonviolation of BIs in the spin-$1$
entangled states can be tested by utilizing the orbital angular-momenta
entanglement.{\textsuperscript{\cite{28,29,32}}}

\part{\protect\large Acknowledgements}

\smallskip This work was supported in part by the National Natural Science
Foundation of China, under Grants No. 11275118, U1330201.

\part{\protect\large Appendix}

Proof of the modified BI $|P_{lc}(ab)-P_{lc}(ac)|\leq 1-P_{lc}(bc)$ for the
entangled state with parallel spin-polarization in terms of classical
statistics following Bell.

The correlation of product expectation-values for measuring two spins
respectively along unit-vector directions $\mathbf{a},$ and $\mathbf{b}$ is
evaluated as
\begin{equation*}
P\left( a,b\right) =\int d\lambda \rho \left( \lambda \right) A\left(
\mathbf{a},\lambda \right) B\left( \mathbf{b},\lambda \right) ,
\end{equation*}%
where $\rho \left( \lambda \right) $ is the probability distribution and $%
\lambda $ is a hidden variable. $A\left( \mathbf{a},\lambda \right) =\pm 1$
and $B\left( \mathbf{b},\lambda \right) =\pm 1$ denote the outcome
expectation-values of measurements for two spins. The measuring outcome
expectation-values of two spins\ are equal $A(\mathbf{a}$,$\lambda )=B\left(
\mathbf{b},\lambda \right) $ for the entangled state Eq.(1) with parallel
spin-polarization. Hence

\begin{equation*}
P\left( a,b\right) =\int d\lambda \rho \left( \lambda \right) A\left(
\mathbf{a},\lambda \right) A\left( \mathbf{b},\lambda \right) ,
\end{equation*}%
and

\ \ \ \ \ \ \ \ \ \ \ \ \ \
\begin{equation*}
p\left( a,b\right) -P\left( a,c\right) =\int d\lambda \rho \left( \lambda
\right) A\left( \mathbf{a},\lambda \right) A\left( \mathbf{b},\lambda
\right) \left[ 1-A\left( \mathbf{b},\lambda \right) A\left( \mathbf{c}%
,\lambda \right) \right] .
\end{equation*}%
It is obviously that
\begin{eqnarray*}
\left\vert P\left( a,b\right) -P\left( a,c\right) \right\vert &\leqslant
&\int d\lambda \rho \left( \lambda \right) \left[ 1-A\left( \mathbf{b}%
,\lambda \right) A\left( \mathbf{c},\lambda \right) \right] \\
&=&1-P\left( b,c\right)
\end{eqnarray*}%
since $\rho $ is a normalized probability distribution $\int d\lambda \rho
\left( \lambda \right) =1$.

\part

\end{document}